\documentclass[epjST,final]{svjour}
\usepackage{graphics}
\usepackage{graphicx}
\usepackage{dcolumn}
\usepackage{bm}
\usepackage{amsmath}
\usepackage{xcolor}
\usepackage{subfig}
\usepackage[normalem]{ulem}

\newcommand{\be}{\begin{equation}}
\newcommand{\ee}{\end{equation}}
\newcommand{\ba}{\begin{align}}
\newcommand{\ea}{\end{align}}

\begin{document}

\title{ Beyond Fowler-Nordheim model: Harmonic generation from metallic nano-structures }
\subtitle{Beyond Fowler-Nordheim model for metal nano-structures}
\author{S. Yusofsani\inst{1} \and M. Kolesik\inst{1}\fnmsep\thanks{\email{kolesik@acms.arizona.edu}} }
\institute{College of Optical Sciences, University of Arizona, Tucson, AZ 85721, U.S.A.}
\abstract{
Metallic structures interacting with electromagnetic fields are known to exhibit
properties similar to those found in atoms and molecules, such as multi-photon and tunnel ionization.
Developing this similarity beyond the electron emission current, we generalize the well-known Fowler-Nordheim
model, and predict heretofore unrecognized source of nonlinear optical response from nano-structures
exposed to illumination with intense optical pulses.
} 
\maketitle
\section{Introduction}

The research into physics of structured surfaces and in particular nano-structures has been the subject
of a growing interest. In this context, one of the richest areas of investigation has been
the non-linear light-matter interactions~\cite{Kruger_2018,AttoNano17,SommerEtal,Kruger_2012} and in particular
the emission of electrons due
to irradiation by high-intensity optical pulses~\cite{nt_twocolor,nt_control,classSHG,Shou21HHG,Moreau_2019,PhysRevX.9.021029}.
Nonlinear optical properties of nano-scale surfaces is another field  which attracts
considerable attention~\cite{RevModPhys.92.025003,khalid2020secondharmonic,respfunct,QCbridge}.

While the relation between the emitted
electron current and the electric field intensity has been studied for almost a century~\cite{ShiDong,FowlerNordheim,OldReview},
recent experiments open a whole new range of opportunities for applications and basic research~\cite{AttoNano17,nt_twocolor,nt_control}.

It has been recognized that the behavior of nano-structures exposed to the external optical pulse exhibits
some important similarities to the dynamics of single atoms in the external field~\cite{Kruger_2012} such as multiphoton
ionization~\cite{VarroEtAl} and strong field ionization~\cite{PhysRevLett.105.257601}. However, so far the
parallels have been recognized mainly for the electrons liberated from the system.

In atomic
and molecular systems there is always a nonlinear polarization response which accompanies the strong-field
ionization~\cite{Couairon2007}.
In fact, whenever the external  field becomes strong enough to cause field-ionization in atomic species,
it also necessarily gives rise to a dipole moment as a result of the deformation of electronic wavefunctions
in bound-states~\cite{SSSIvsMESA}. As a result, the ionization and the nonlinear polarization of atomic and molecular species
exposed to strong electric fields are closely connected~\cite{kolesik2019universal}. They together give rise to physics driven by
both bound and freed electrons and their interplay governs many phenomena in modern nonlinear optics, including
optical filamentation~\cite{Milchberg:14} and long-distance localized pulse propagation~\cite{MegaF}.
It is fair to say that whenever ionization
by strong electric field can be detected, the nonlinear polarization response is already so strong that it
must not be neglected. This begs a question if such a connection should be made for nano-structures.

In this paper we take the atom-nanostructure analogy one step further and find that a strong polarization
response exists along the tunneling electrons.  We aim to show that these two aspects are intimately related
and wherever there is a tunneling current, one should also look for a surface polarization accompanying electron emission.
Our`s is a conceptual study based on a simple model which is in fact exactly solvable. While it is certainly
too simple to hope for a {\em quantitative} description of any real system,  it can be studied in detail analytically
and numerically and allows us to establish a direct connection between the polarization and the current
facets of the optical response of metallic nanostructures.

One of the widely used results in the context of field-induced current caused by external electric field is
the Fowler Nordheim (F-N) relation~\cite{FowlerNordheim,OldReview}. The F-N formula relates the intensity of external
field to the tunneling current and the treatment is based on a one-dimensional single-particle model.
In the present work, we first re-create the F-N scenario and reformulate it using a non-Hermitian formalism.
Deviating from the standard F-N approach, we use the quantum resonance states rather than the more conventional
scattering states. Having verified our approach via comparison with the F-N model, we then identify the nonlinear
polarization produced at the surface of the sample. In order to estimate the strength of this interaction,
we make a comparison in terms of per-atom nonlinear dipole, and find it comparable to that induced by strong fields
in the most nonlinear noble-gas atoms. Finally we discuss possible experimental signatures, specifically in
the form of higher-harmonic radiation.

\section{Model}

The physical model used in this work is essentially the same as what underlines the well-known Fowler-Nordheim description
of electron emission from a metal structure exposed to an electric field. It is assumed here that electrons in the
conduction band can  be approximated in terms of the free-electron model, and on the outside of the sample
there is an electric field that pulls electrons away from the sample surface. In order to account for the three-dimensional
density of electronic states, we shall later consider a 3D volume, but in order to keep notation simple, we
can start with a one-dimensional description.

The differential expression associated with the action of the Hamiltonian is split into domain of
$-L\le x \le 0$ which represents the conduction band  and the exterior $0<x<\infty$ where a homogeneous external
field of strength $F$ is applied:
\begin{align}
  H &= -\frac{1}{2} \frac{d^2}{d x^2}  - V_0 \ \ \ \  \text{for} \ \ \ -L \le x \le 0  \nonumber \\
  H &= -\frac{1}{2} \frac{d^2}{d x^2}  - s F x \ \ \ \  \text{for} \ \ \  x > 0 \ .
\label{eqn:hamiltonian}
\end{align}
For the sake of notation, we will assume that quantity $F$ is always positive, and parameter $s$ above is used
to indicate whether this field tends to pull particles away from the sample surface ($s>0$) or it wants to
push the electrons back into the metal ($s<0$).

To select the domain of the Hamiltonian, we ask for the following boundary and continuity conditions to be satisfied: 
\be
\psi(-L) = 0 \ , \ \psi(0^-) = \psi(0^+) \ , \ \psi'(0^-) = \psi'(0^+) \ .
\ee

Parameter $V_0$ represents the depth of the conduction band, and we will use $\phi$ to denote the workfunction
of the material. While $L$, the length of the metallic sample is macroscopic,  we carry out all calculations
for a finite $L$, and limit $L\to\infty$ will be taken for observable quantities. This model is exactly solvable
and all quantities can be calculated analytically for arbitrary sample length, but the macroscopic limit will result in greatly
simplified expressions. 

Later, when we need to account for the three-dimensional nature of the electron ensemble, the
energy eigen functions will take the form
\be
\psi_W(x) \exp[ i k_y y + i k_z z]
\ee
with the first factor being the eigenfunction of (\ref{eqn:hamiltonian}) corresponding to energy $W$, and where $k_{y,z}$ stand
for the momenta in the transverse directions. For the sake of simplicity we will treat the
electrons at the zero temperature, with all states below the Fermi energy being occupied. We will use $k_F$ to denote
the Fermi velocity in the conduction band.

\section{Strong-field electron emission}

As a first step, we will calculate the rate of emission of particles tunneling from the conduction band into
the vacuum when the external field points away from the surface, i.e. for $s>0$. Here we employ a non-Hermitian approach
utilizing Stark resonant states. The rationale for this deviation from the standard quantum description is that
it allows one to go beyond the emitted current and also calculate the concomitant nonlinear induced surface dipole
moment which gives rise to up-converted electromagnetic radiation. The following two subsections aim to demonstrate that
the non-Hermitian approach can reproduce the results of the Fowler-Nordheim model, and we do this to validate our
approach.

\subsection{Non-Hermitian treatment}

Stark resonances in the context of this work are the energy eigenstates of the differential equation (\ref{eqn:hamiltonian})
which obey the so-called Siegert outgoing-wave boundary conditions for $x\to\infty$~\cite{siegert}. For the outside component of the
wavefunction they can be found in terms of the solutions to Airy equation like so
\begin{align}
  \psi_W(x) &= \sin(k_W (L+x)) \text{Ci}^+(\alpha (\phantom{+}0 + W/F)) \ \  x<0 \nonumber \\
  \psi_W(x) &= \sin(k_W (L+0)) \text{Ci}^+(\alpha (+x + W/F)) \ \  x>0 
\end{align}
where
\be
\alpha = -(2 F)^{1/3} \ , \ k_W = \sqrt{2 (V_0 + W)}
\ee
and $\text{Ci}^+ = \text{Bi} + i \text{Ai}$ is the combination of Airy functions that behaves as the outgoing wave
at large distances. In order to satisfy the requirement of the continuous derivative, complex-valued energy $W$
must be chosen as to satisfy this eigenvalue equation:
\be
k_W \cos(k_W L) \text{Ci}^+(\alpha W/F) =  \alpha \sin(k_W L) \text{Ci}^{+'}(\alpha W/F) \ .
\ee
We are interested in the solutions for which the real part of $W$ is between the bottom of the conduction band
and below the Fermi energy, i.e. we seek
\be
-V_0 < \text{Re}\{W\} < -\phi \ .
\ee
At the same time, we consider a large $L$, and the trig functions will localize each $\text{Re}\{W\}$ in an interval
of size that scales with $1/L$. So one can expect to obtain a quasi-continuum of solutions, and we turn our attention
to their imaginary parts. Because of the density of solutions, we may consider the imaginary part of $W$ as being controlled
by the real part.

For values of $F$ which correspond to the range of the laser intensities typically used in experimental setups,
the  $\text{Ci}$ in the eigenvalue equation is dominated by $\text{Bi}$ which is exponentially larger
than the $\text{Ai}$ contribution, as can be verified with the help of their asymptotic forms,
\be
\exp[+2 \sqrt{2} |W|^{3/2} /(3 F)] \sim \text{Bi}(\alpha W/F) \gg \text{Ai}(\alpha W/F) \sim  \exp[-2 \sqrt{2} |W|^{3/2} /(3 F)] \ .
\ee
This invites us to treat the $\text{Ai}$ part of the eigenvalue equation as a small term and look for the
solution in the form of $W = W(F) + \delta W(F) + \ldots$  where the
dominant real-valued $W(F)$ obeys the equation  obtained via replacement $\text{Ci}^+\to\text{Bi}$:
\be
k_W \cos(k_W L) \text{Bi}(\alpha W/F) =  \alpha \sin(k_W L) \text{Bi}'(\alpha W/F) \ .
\ee
Then the first correction $\delta W(F)$ is obtained by inserting $W(F)$ in the small perturbation,
and expanding the dominant part in $\delta W$. As expected, the correction turns out to be purely imaginary:
\be
\delta W(F) = \frac{2 i \alpha (V_0+W)}{\pi L \left(2  (V_0+W) \text{Bi}^2 + \alpha^2 \text{Bi}'^2   \right)}
\approx
-i \frac{\sqrt{-2 W} (V_0 + W)}{L V_0} \exp[-\frac{4 \sqrt{2} |W|^\frac{3}{2} }{3 F}]
\label{eqn:imagW}
\ee
To simplify our notation, we used here, and/or will use in what follows
\be
\text{Bi} \equiv \text{Bi}(\alpha W/F) \ , \ \text{Bi}' \equiv \text{Bi}'(\alpha W/F)
\ , \ 
\text{Ai} \equiv \text{Ai}(\alpha W/F) \ , \ \text{Ai}' \equiv \text{Ai}'(\alpha W/F) \ .
\ee
As a result we obtain the ionization rate of the state with the real part energy $W$
\be
\Gamma(F) = -2 \text{Im}\{\delta W(F)\} =  \frac{4 \alpha (V_0+W)}{\pi L \left(2  (V_0+W) \text{Bi}^2 + \alpha^2 \text{Bi}'^2   \right)}\ .
\ee
   The macroscopic current is obtained by summing up the contributions from all occupied states in the
conduction band, and in that process dependence on $L$ disappears as should be expected.
We will do this explicitly for the polarization components of the response to field $F$ in Section 4.
Here we want to concentrate on comparison with the F-N model. The asymptotic form in (\ref{eqn:imagW})
already indicates that the low-field behavior is in fact the same as in the F-N model. In the following
section we show that the current aspect of our treatment is in fact equivalent to that of the F-N framework.

\subsection{Comparison with Fowler-Nordheim treatment}

For the sake of completeness the F-N framework is summarized next.
It utilizes the scattering states, which one can choose to parameterize as follows
\begin{align}
  \psi_W(x) &= \frac{1}{N_s}\exp(+i k_W x) + \frac{1}{N_s} R(W) \exp(-i k_W x) \ \  x<0 \nonumber \\
  \psi_W(x) &= \frac{1}{N_s} T(W) \text{Ci}^+(\alpha (+x + W/F)) \ \  x>0 \ ,
\end{align}
where $R(W)$ and $T(W)$ are fixed so that the wavefunction is smoothly continuous at $x=0$.
The normalization factor is in general complicated, but to the leading order in size of the system $L$
it would be $N_s \sim (2 L)^{1/2}$.
The probability current leaking to  the positive $x$-axis can be expressed as
\be
J(W) = \frac{1}{N_s^2} k_W ( 1 - |R(W)|^2) \approx
\frac{4 \alpha  (V_0+W)}{\pi L \left(2  (V_0+W) \text{Bi}^2 + \alpha^2 \text{Bi}'^2   \right)} \ .
\ee
On the left is an exact expression, and when evaluated for $F$ so weak that
$\text{Bi} \gg  \text{Ai}$
one obtains the expression on the right. 
This quantity should be compared to the Stark-resonance decay rate of $2 \text{Im}\{\delta W(F)\}$.
Thus, the rate of leakage from any given state is the same in the Stark and F-N models as long as one does not commit to unnecessary
approximations in F-N treatment. This equivalency is a validation of our approach on the current side, this
giving us a license to use the non-Hermitian treatment to go beyond F-N and obtain the polarization response in what follows.

{
  It is worth mentioning that there are many ways in which the F-N can be improved e.g. by considering a general
  potential barrier~\cite{doi:10.1063/1.2354582,doi:10.1063/1.2937077} or by taking into account the effect of image charge
  induced by electron on the surface of the metal~\cite{NordheimImage,SchottkeyImage} or by including the local field enhancement,
  and the effect of finite temperature~\cite{PhysRev.113.110,RevModPhys.45.487}. While many of such improvements can be also applied
  in our treatment, it is not our aim for this work. Rather, we prefer to keep the model as simple as possible in order to
  emphasize that the appearance of the nonlinear polarization, which is discussed in the next section, is a universal effect
  reflecting the fact that the two kinds of response are in fact different sides of the same dynamics.
}

\section{Nonlinear induced polarization}

Now that we have successfully reproduced the F-N per-state current in the non-Hermitian formulation, we can show another important aspect of this formalism,
namely the calculation of  the dipole moment and eventually the surface polarization per atom for the surface of the nano-structure irradiated by the
external field. This is an experimentally testable prediction which could not be made using the scattering states utilized by  the F-N model.

As we will see in the coming sections, the resulting dipole moment shows a highly non-linear behavior with respect to the external field.
The physical reason can be inferred from the fact that the electric field does not penetrate easily beneath the metallic surface.
The formal, or mathematical reason for such behavior is that the calculation of the dipole moment for positive and negative fields
must be done separately because the asymptotic behavior of the dominant Airy function in the wavefunction changes when $F$ changes sign.

\subsection{ Dipole moment calculation for positive field}

Instead of calculating the expectation value of the dipole moment by integration
over wavefunction products, we can use the relation that connects the dipole moment to the
rate of change of the energy eigenvalue with respect to the field strength~\cite{brown2015properties}, namely
\be
M = -\partial_F W(F) \ .
\ee
The direct calculation of the dipole moment is in principle possible, but it results in rather unwieldy expressions.
In contrast, the above equation can be used in conjunction with 
the field-differentiated eigenvalue equation,
\be
\partial_F \left[  k_W \cos(k_W L) \text{Bi}^+(\alpha W/F) -  \alpha \sin(k_W L) \text{Bi}^{+'}(\alpha W/F) \right] = 0 \ ,
\ee
to obtain the quantity of interest. This calculation simplifies considerably when retaining only the leading order
in large $L$ and the result reduces to
\be
M(F) = \frac{ 2(V_0+W)\left[ 4 W^2 \text{Bi}^2 - 2^{1/3} F^{4/3}\text{Bi} \text{Bi}' + 2^{5/3}F^{2/3} \text{Bi}'^2 \right]  }{3 F^2 \left( 2(V_0+W) \text{Bi}^2 + (2 F)^{2/3} \text{Bi}'^2 \right) L}  \ . 
\label{eqn:MFpos}
\ee
Of course, the constant ``background'' value of 
$M(0) = (W_0+V_0)/(4LV_0W_0)$
is irrelevant for the interaction with light since the relevant physical quantity is the change in the dipole moment,
and we shall subtract $M(0)$ in the following.

We have thus derived a contribution to the induced polarization of the sample which originates in a single quantum state.
This constitutes the counterpart of the per-state current as it was derived either in the original F-N framework or in 
our non-Hermitian approach. It becomes clear that the two aspects, namely the polarization and current are expressions of the
same mechanism in which the single-particle state adopts to the value of the external field.
To complete the picture we must turn to the remaining half of the calculation, when the field pushes the particle from the
exterior toward the sample surface.

\subsection{ Dipole moment calculation for negative field}

\begin{figure}[h]\sidecaption
  \centerline{\includegraphics[clip,width=0.4\textwidth]{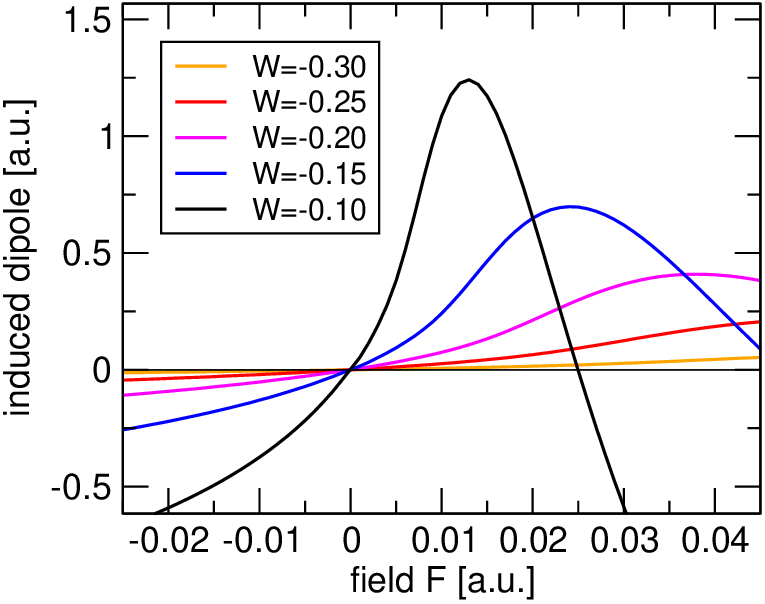}}
\caption{Induced dipole versus field strength for several states with different energies.
  The shape of these curves for $F>0$ closely resembles the induced dipole moment in noble-gas atoms.
  However, the peaks occur at significantly lower field intensities because the
  relevant energy scale is much smaller than ionization potentials of atoms.
   }
\label{fig:indd}  
\end{figure}

When the field direction is into the structure ($s<0$), the outside solution is best expressed in terms of
Airy function $\text{Ai}$:
\begin{align}
  \psi_W(x) &= \sin(k_W (L+x)) \text{Ai}(\alpha (\phantom{+}0 + W/F)) \ \  x<0 \nonumber \\
  \psi_W(x) &= \sin(k_W (L+0)) \text{Ai}(\alpha (+x + W/F)) \ \  x>0 \ .
\end{align}
In this case, the energy spectrum is purely discrete (and real, of course), and the eigenvalue equation is obtained as
\be
k_W \cos(k_W L) \text{Ai}(\alpha W/F) = -\alpha \sin(k_W L) \text{Ai}^{'}(\alpha W/F) \ .
\ee
As in the case of positive field, the induced dipole moment can be obtained from the differentiated
eigenvalue equation which leads, again including solely the leading term in $1/L$, to
\be
M(F) = \frac{ 2(V_0+W)\left[ 4 W^2 \text{Ai}^2 - 2^{1/3} F^{4/3}\text{Ai} \text{Ai}' + 2^{5/3}F^{2/3} \text{Ai}'^2 \right]  }{3 F^2 \left( 2(V_0+W) \text{Ai}^2 + (2 F)^{2/3} \text{Ai}'^2 \right) L} \ .
\label{eqn:MFneg}
\ee
The  results for positive and negative fields are put together and illustrated in the 
Fig.~\ref{fig:indd}, depicting the resulting microscopic induced dipole moment
for several energies.

The behavior for very strong fields reflects the fact that the classical exit point from
   the potential barrier approaches the boundary of the system, and thus causes the wavefunction
   to decay faster as it leaks out. It should be noted that a similar behavior can be found in
   long-lived unstable Stark resonances in noble-gas atoms~\cite{kolesik2019universal}. However, the region in which the dipole curve
   starts to bend downwards is impossible to reach in gases due to propagation effects caused by
   the concomitant ionization. It is conceivable that a similar ``screening'' can occur in strongly
   driven metallic tips.

   One can also see that for the range of energies in the
   vicinity of Fermi-level typical for metals, the response changes fast with the state energy.
   These are the states that contribute the most to the macroscopic response.
  The asymmetric and non-linear shape of these curves indicates that the induced dipole driven by
   a harmonic field with certain frequency will radiate strong higher-harmonic signals.

\subsection{Macroscopic response}

To calculate the macroscopic response of the nano-tip to the irradiation by intense light, we need to add up all
microscopic contributions to the induced dipole moment. Using the free-electron model for the conduction band
at zero temperature, we  integrate over states with the three-dimensional momenta less than
the Fermi velocity $k_F$. The latter is related to the electron density $n$ (in the conduction band)
via $k_F^3 = 3 \pi^2 n $. 

To obtain the field-induced dipole moment, we sum up all contributions expressed in (\ref{eqn:MFpos},\ref{eqn:MFneg}) for the populated states:
\be
P(F) = 2\times 2 \pi \int_{-k_F}^{+k_F} M(k_W,F)  \int_0^{\sqrt{k_F^2-k_W^2}} k_\perp dk_\perp \frac{L^3}{(2\pi)^3} dk_W \ .
\ee
The above expression makes it evident 
that the observed dipole moment is proportional to the surface area of the sample, and it therefore originates in the space
adjacent to the nanostructure surface. Before continuing, let us obtain an order of magnitude estimate of the dipole moment
per atom residing on the surface. The surface density of the dipole is
\be
p(F) = P(F)/L^2 = 
 4 \int_0^{k_F} L M(k_W,F) (k_F^2 - k_W^2)  \frac{1}{(2\pi)^2} dk_W \sim k_F^3 \sim n
 \ee
Quantity $L M$ turns out to be of the order of unity so the estimate for
dipole moment per atom at the surface, with $a_L$ standing for the lattice constant, is
\be
p_a(F) = p(F)\times\text{area per atom} \sim n a_L^2  \sim 5\times 10^{-2} \text{a.u.}
\ee
This is a very rough estimate, but it is in the same range as the nonlinear dipole per noble-gas atom. 
It is an indication that the nonlinear response, while localized at the nano-structure surface,
can be significant.

For a more quantitative picture, we proceed to evaluate the nonlinear response for a few examples
motivated by the metallic nano-tips. Assuming that the material is characterized in terms of its work
function $\phi$, together with the Fermi velocity $k_F$, we can obtain parameter
$V_0= \phi + \frac{1}{2} k_F^2 $ and proceed to integrate numerically:
\be
p(F) =   \frac{-2}{\pi^2} \int_{-V_0}^{-\phi} L M(W,F) \frac{ (\phi + W)}{\sqrt{2 (V_0+W)}}  dW  \ .
\ee
The induced dipole moment turns out to be mainly controlled by the states that are close to the
Fermi level, as intuitively expected. The overall shape of the response to the external field
resembles the nonlinear dipole moment induced in atoms by optical fields, at least for the
positive field values. For negative fields, the shape of the response curve is flatter, thus giving
rise to an overall asymmetric response causing even-harmonic generation.

\begin{figure}[!h]\sidecaption
    \centerline{\includegraphics[clip,width=0.45\textwidth]{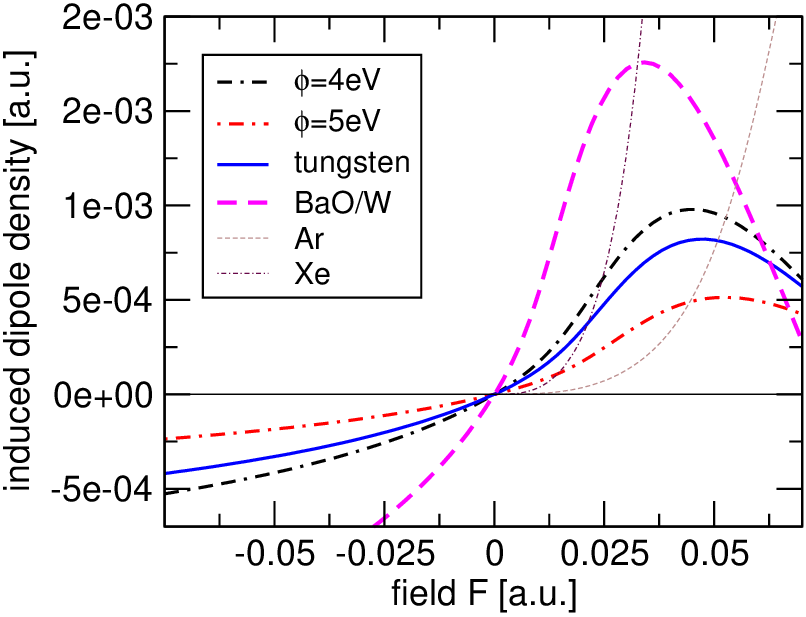}}
\caption{ Induced dipole moment density obtained for workfunction values $\phi$ representing the range typical
  of metals. For comparison, thin lines represent the response of noble-gas atoms scaled to
  surface atom-density of tungsten. 
}
\label{fig:macroD}       
\end{figure}

These properties are illustrated in Fig.~\ref{fig:macroD}.
   Nonlinearity of the curve implies strong harmonic generation, and its asymmetry means that 
   even-order harmonics will also be generated.
   For comparison, nonlinear response of noble gas atoms, scaled to the surface atom-density of tungsten is
   also shown to demonstrate that the per-atom response on tungsten is comparable to that of Argon,\cite{bahl2017nonlinear} and
   for Barium-Oxide coated tungsten it gets as strong as that of the most nonlinear noble gas of
   Xenon~\cite{tolliver2020nonlinearity}.

\section{ Nonlinear optical response: Experimental signatures }

Manifestations of the nonlinear dipole density induced on the surface of a metallic nano-structure
will obviously depend not only on the material but also on the geometry of the surface and  on the
parameters of excitation by optical pulse. Here we wish to consider harmonic generation as one possible effect
that could be utilized as means of detection of the nonlinear response mechanism put forward in this work.
We envision a situation in which a large collection of nano-tip structures~\cite{nt_carpets,HanEtal} is irradiated by  an
optical pulse, giving rise to a coherent array of nonlinear oscillating dipoles. The spectral
content of the re-radiated field will reflect the properties of the individual structure, and this
we look at next.

What sets the surface nonlinear response apart from the harmonic generation due to e.g. Kerr effect is the
difference in the relative strength of different harmonic bands. In the usual perturbative regime of the
third-order nonlinearity, third harmonic radiation is generated first and then higher-orders occur via 
cascade. The result is a conversion efficiency which decreases exponentially with the order, with an
order-to-order power-drop of two to three orders of magnitude.
In contrast, the asymmetric and nonlinear shape of the induced surface dipole density gives rise
to all harmonic orders at the same time, and this includes both even and odd harmonic bands.
As we will see, this response is highly nonlinear and non-perturbative in that higher
order harmonic radiation generated from a nano-structure are comparable in strength to that of the lower orders. 
In a sense, the behavior is akin to high-harmonic generation~\cite{HHGplateau} in gases when a plateau can form in which
different orders carry almost the same power. However, this is something rarely occurring at lower
harmonic orders, and can thus serve as an experimental signature of the effect described here.  

We assume that the frequency of the field is in the optical range, and therefore much smaller than the
atomic-scale frequencies. As a consequence the response is adiabatically following the driving field.
To illustrate spectral properties of the nonlinear response, we take an example of a  pulsed driving electric
field, specified as a function of time in the form:
\be
A(t) = F \sin(\omega_0 t) e^{-t^2/\tau^2}
\ee
Using the result of previous sections, the induced nonlinear polarization $P(t)$ of the surface is 
a map of the excitation field via the nonlinear dipole function. We are mainly interested in its
spectrum $\hat P(\omega)$:
\be
P(t) = p(A(t))  \ \ \ , \ \ \ \hat P(\omega) = \int e^{+i \omega t} p(  F\sin(\omega_0 t) e^{-t^2/\tau^2} ) dt \ ,
\ee
in which we aim to compare power carried by different harmonic bands.

\begin{figure}[!h]
  \centerline{\includegraphics[clip,width=0.45\textwidth]{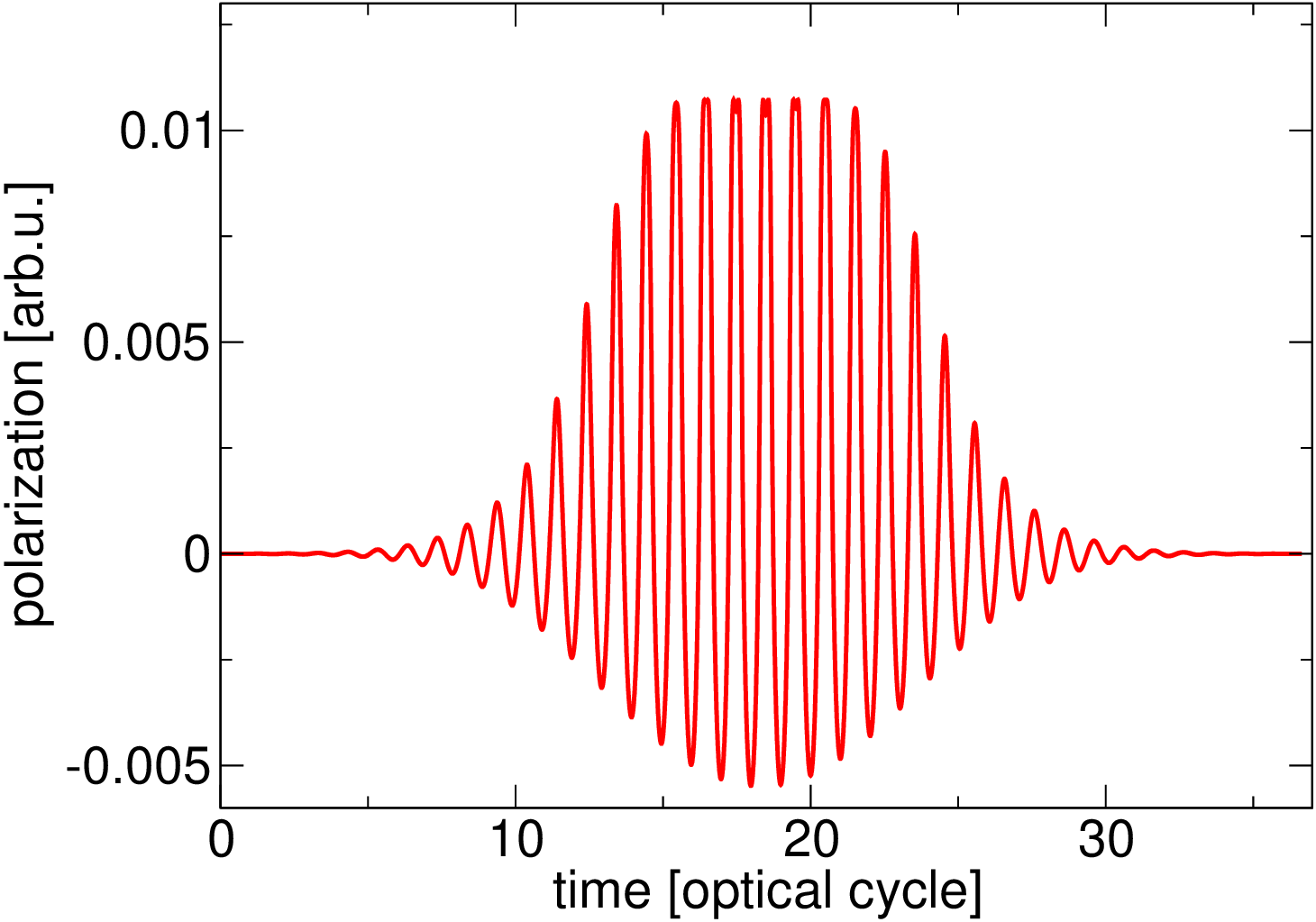}
              \includegraphics[clip,width=0.45\textwidth]{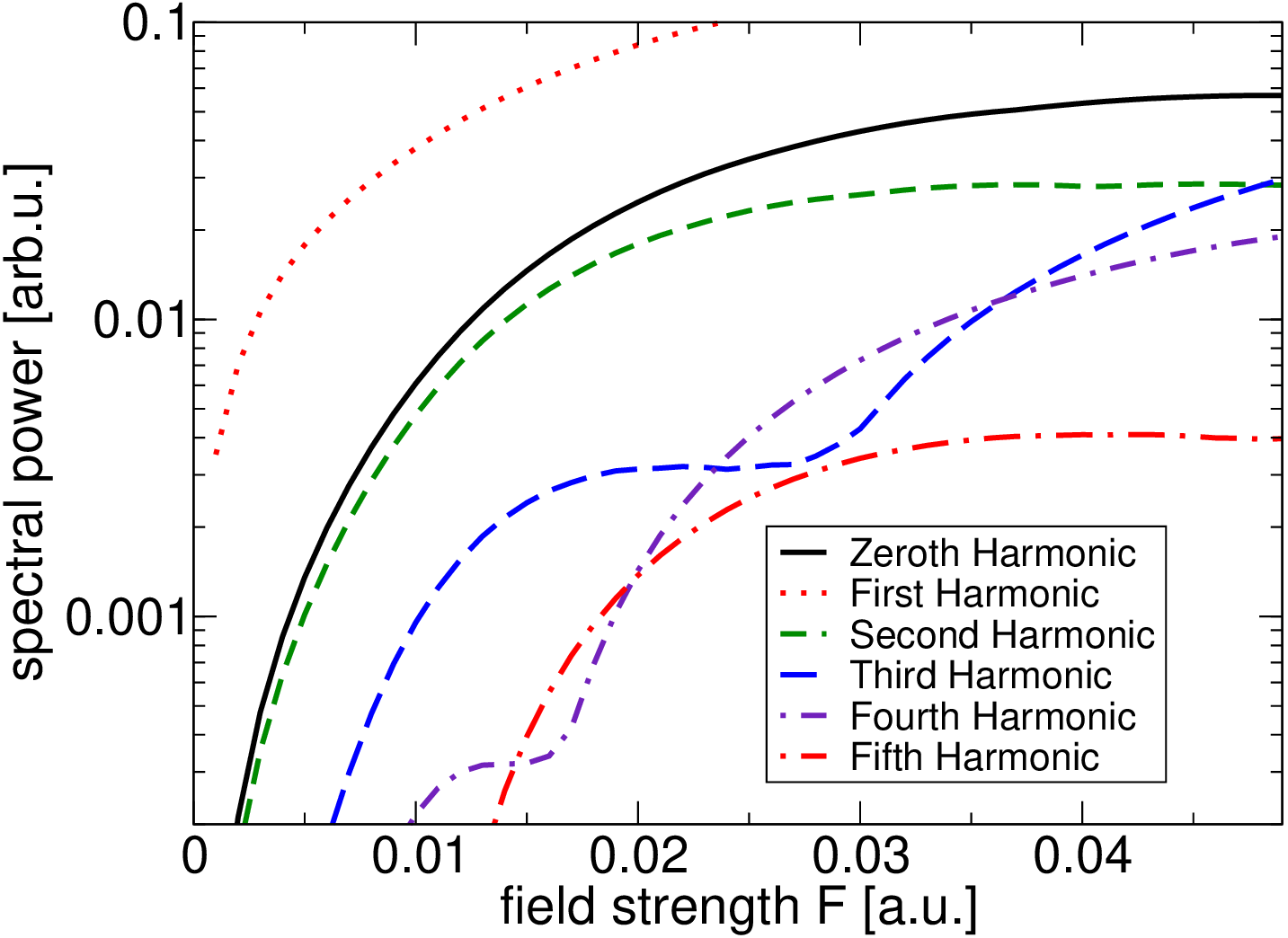}}
  \caption{Harmonic generation due to nonlinear polarization on surface of tungsten.
    Left: A sample of polarization induced by a driving pulse with amplitude $F=0.04$a.u.
    Right: Spectral power integrated over harmonic-frequency bands exhibits relatively
    strong contribution of higher harmonics.
  \label{fig:HG}
  }
\end{figure}

Figure~\ref{fig:HG} illustrates the non-perturbative nature of the nonlinear dipole induced
on the surface of tungsten taken as an example of a material often utilized to manufacture metallic
nano-tips. For this illustration, it was assumed that a nano-tip was irradiated by an optical pulse,
and we calculated the surface polarization response. This is shown in the left panel of the figure.
An interesting signature is a well-developed asymmetry of the polarization curve, which signifies
strong second-harmonic generation as well as an optical rectification signal in the ``zero-harmonic'' 
frequency range. One can also clearly see that different functional shape occurs around the peaks,
and this reflects the fact that higher harmonic radiation will also be generated. The panel on the
right hand side illustrates the power integrated in each harmonic frequency band and shows
several of these bands as functions of the driving field amplitude.

\begin{figure}[!h]
    \centerline{\includegraphics[clip,width=0.45\textwidth]{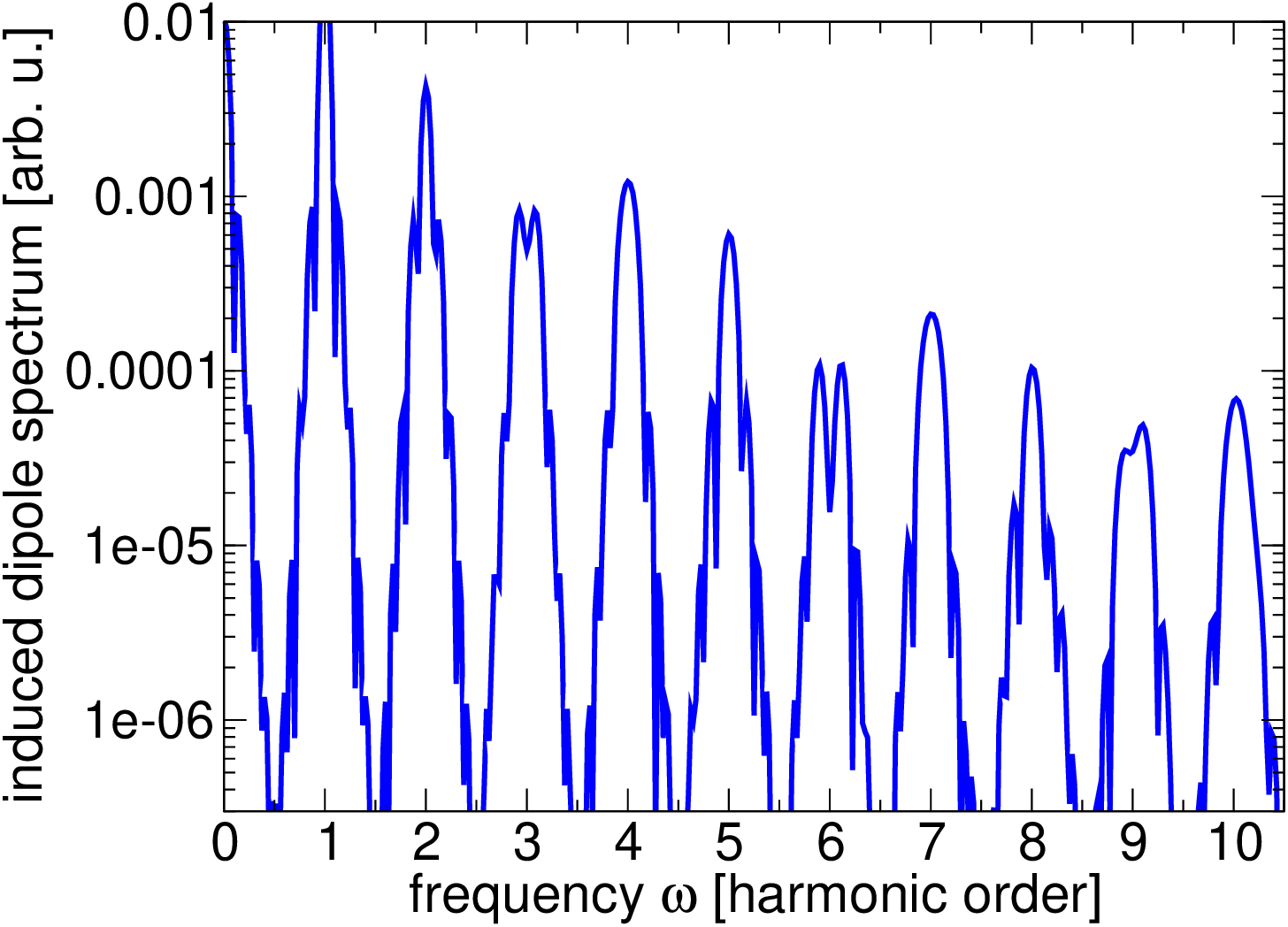}}
  \caption{High-harmonic spectral content of the nonlinear induced dipole for a strong driving field F=0.08a.u.
    While in general this mechanism does not give rise to a plateau like in the true high-harmonic generation,
    a large number of harmonic bands can exhibit power in a narrow range of just a few orders of magnitude.
  \label{fig:HHG}
  }
\end{figure}

The important property that reflects the highly nonlinear nature of the response is that
several harmonic frequencies exhibit power levels which are quite comparable, being
all within a one to two orders of magnitude range when the irradiating field is strong,
which is illustrated in Fig.\ref{fig:HHG}
It can even happen that a higher harmonic becomes stronger than its lower-order counterpart.
The non-monotonic increase with the driving field amplitude is another signature that seems
universal although the precise shape depends on the material (i.e. mainly on its work function)

\section{Conclusion}

The main outcome of this work is
the generalization of the Fowler-Nordheim model for the electron emission
from a metal surface under the influence of external electric field.
We have reformulated this well-known approach
in the non-Hermitian language of metastable states, and this allowed us to reveal that concomitant with the emission current,
there occurs a highly nonlinear induced dipole moment density at the surface of the nano-structure.

While it has been recognized for some time that the response of metallic nano-tips and nano-structures in general
bears some similarity to atoms driven by strong electric fields, notable in the form of tunnel ionization and
multi-photon ionization, the question was if such similarities can be identified in other aspects of their dynamics. This was one of our
motivations, and indeed our finding extends the atom-nanostructure analogy from the emitted current to include the
induced dipole moment as well. In a close analogy to the atomic species responding to the electric fields via emitted electron current
together with the nonlinear polarization, we present an exactly solvable model suggesting that whenever the field
is strong enough to drive the electron emission, there should exist an induced dipole moment density localized at the
metallic surface. In optical pulses with wavelength in the near infrared and longer, the response can be considered
adiabatic, i.e. slaved to the time-dependent electric field resulting in a radiation-source term for Maxwell equations.

The exact solution for a static field case allows one to estimate the strength of this nonlinear response. It turns out that
in terms of the induced dipole moment per atom on the surface, the response is roughly as strong or even stronger than that
observed in the noble gases like Argon and Xenon~\cite{bahl2017nonlinear,tolliver2020nonlinearity}. This finding suggest that the effect should be possible to detect
in future experiments. Indeed, radiation even from individual nanostructures can be detected (e.g.~\cite{respfunct,PhysRevLett.98.043907}), the
question is how to differentiate the proposed effect from other nonlinear mechanisms, including the classical
second-harmonic generation from surfaces, and high-harmonic generation enhanced by nano-structured surfaces (e.g. ~\cite{LiuEnhance}).

Perhaps most promising avenue presents itself in frequency conversion, for example by irradiating carpets~\cite{nt_carpets,sitiparray} of nano-tips
with femtosecond pulses of high intensity. While it is difficult to estimate the total converted power, 
we have investigated possible experimental signatures in the spectra of the polarization driven by pulsed excitation.
The expected harmonic radiation can be characterized by power which decreases very slowly with the harmonic order.
Besides even and odd harmonics, optical rectification is also expected.
Together with the non-monotonic growth of the harmonic power with the driving field amplitude, these signatures
can potentially identify the specific mechanism we put forward in this work. 

Another potential pathway for experimental investigations could borrow the idea of exposing atoms to the enhanced near-field
in the vicinity of a nano-structure~\cite{KimEtAl,Husakou,PhysRevA.87.041402,GaetaNano}.
Atoms or molecules in a close proximity of a light-driven nano-tip would experience high-harmonic field produced by the surface
which in turn could result in the species excitation and subsequent radiation.

In conclusion, we have presented a generalization of a well-known and often utilized model, and this made it possible to
extend the analogy between the nonlinear response of atoms on one side and nano-structures on the other beyond
the emission of electrons, now including also the nonlinear polarization.

\noindent
{\bf Acknowledgments}\\
This material is based upon work supported by the Air Force Office of Scientific Research under
award number FA9550-18-1-0183.

\end{document}